%%%%%%%%%%%%%%%%%%%%%%%%%%%%%%%%%%%%%%%%%%%%%%%%%%%%%%%%%%%%%%%%%%%%%%%%%
\magnification=\magstep1
\baselineskip=12pt

\font\eightit=cmti8
\font\eightpoint=cmr8

\def\lanbox{\hbox{$\, \vrule height 0.25cm width 0.25cm depth
0.01cm \,$}}

\def\d{{\rm d}}

%%%%%%%%%%%%%%%%%%%%%%%%%%%%

\hbox to \hsize{\hfil\eightit PEEHML 5 Nov, 1998} \vglue1truein
%%%%%%%%%%%%%%%%%%%%%%%%%%%%%%%%%%%%%%%%%%%%%%%%%%%%%%%%%%%%%%%%%%%%%
 \centerline{\bf OPTIMAL EIGENVALUES FOR SOME LAPLACIANS}
 \centerline{\bf AND SCHR\"ODINGER OPERATORS}
 \centerline{\bf DEPENDING ON CURVATURE}

%%%%%%%%%%%%%%%%%%%%%%%%%%%%%%%%%%%%%%%%%%%%%%%%%%%%%%%%%%%%%%%%%%%%%
\bigskip
\bigskip
\bigskip

{\baselineskip=2.5ex
\vfootnote{}{\eightpoint
\noindent\copyright 1998 by the authors.
Reproduction of this article, in its entirety, by any means is permitted
for non--commercial purposes.}}
%%XX The publisher will probably remove this

{\baselineskip = 12pt
\halign{\qquad#\hfil\qquad\qquad\quad\hfil&#\hfil\cr
{\bf Pavel Exner}\footnote{$^*$}{Work supported by
GA AS No.1048801} &
{\bf Evans M. Harrell} \footnote {$^{**}$}{Work supported by N.S.F. grant
DMS-9622730}  and {\bf Michael Loss}\footnote{$^{+}$}{Work supported by
N.S.F. grant DMS--9500840}\cr
&\cr
Nuclear Physics Inst. & School of Mathematics\cr
Academy of Sciences & Georgia Tech\cr
25068 Rez --Prague & Atlanta, GA 30332-0160, \cr
Czech Republic & USA \cr
exner@ujf.cas.cz & harrell@math.gatech.edu \    loss@math.gatech.edu \cr}}
\bigskip

\vskip .7 true in

\centerline{\bf Abstract}
\smallskip

\smallskip
{\rightskip=4pc

We consider Laplace operators and Schr\"odinger operators with potentials
containing curvature on certain regions of nontrivial topology, especially
closed curves, annular domains, and shells.  Dirichlet boundary conditions
are imposed on any boundaries.  Under suitable assumptions
we prove that the fundamental eigenvalue is maximized when the geometry
is round.

We also comment on the use of coordinate transformations for these
operators and mention some open problems.

\par}

\noindent

{\vglue 0.5cm}

\vfill\eject
%%%%%%%%%%%%%%%%%%%%%%%%%%%%%%%%%%%%%%%
\noindent{\bf I. Introduction}
%%%%%%%%%%%%%%%%%%%%%%%%%%%%%%%%%%%%%%%
\medskip

In this article we present some theorems in optimal spectral geometry
which were suggested by situations where the physics is strongly controlled
by the geometry of an interface, i.e., a lower-dimensional structure.  Two
specific interests are electrical properties of nanoscale structures (quantum
wires, waveguides, and resonators) [DuEx], and the slow evolution of
interfaces in reaction-diffusion systems such as Allen-Cahn [AlFu].

By ``optimal spectral geometry'' we mean the determination of the
geometry which maximizes or minimizes a certain eigenvalue of a
differential operator.  The archetypal result of this genre is the
Faber--Krahn theorem [Fab] [Kra]:

1.  If one considers the Dirichlet problem for the Laplacian on a region of
specified volume (or area, in two dimensions), then it is the ball (disk)
which uniquely minimizes the fundamental eigenvalue.

Other well-known theorems of optimal spectral geometry include:

2.  (Weinberger [Wei]).  If one considers the Neumann problem for the
Laplacian on a region of specified volume, then the fundamental eigenvalue
is trivially 0.  The first positive eigenvalue is uniquely {\it max}imized by
the ball.

3.  (Hersch [Her1]).  If one considers the Laplacian on closed, simply
connected surfaces of specified area embedded in $R^3$, then, as in the
previous situation, the fundamental eigenvalue is trivially 0.  The first
positive eigenvalue is then uniquely maximized by the sphere.

These theorems are sometimes termed ``isoperimetric'' in analogy
with the classical isoperimetric theorem whereby the ball
minimizes the surface area given a fixed volume.  The
simplicity of the optimizers in these situations may convey a misleading
idea of the subtlety of the analysis involved.  They involve more than
simple rearrangement of the energy functional, and
indeed the optimizers of higher eigenvalues are not so easy to
characterize or even discover.  Moreover, the theorem of Hersch is not
true in all dimensions.  For reviews of this subject,
consult [AsBe], [Ban], and [Her2].

An exception to the statement about higher eigenvalues, however
is the recent result in [HaLo], that the {\it second} eigenvalue
of the Laplace operator penalized by the square of the mean curvature
is uniquely maximized by the sphere.  This result holds in any dimension.
Interestingly, in two dimensions analogous facts can be proved,
using Hersch's technique of conformal transplantation, for
the second eigenvalue of the Laplace operator penalized by
a substantially larger family of potentials quadratic in curvature (see [Har]).

In this article we shall present some new theorems where the fundamental
eigenvalue is optimized by round geometry, and we attempt to shed
light on the role of curvature in the spectra of Laplace and Schr\"odinger
operators.  In Section II, we consider the Laplacian on certain
non--simply--connected domains and show that, under some circumstances,
the optimization of the fundamental eigenvalue contrasts with the
Faber--Krahn theorem.  In Section III we review a
transformation which has long been used to understand connections
between curvature and spectra for quantum wires and waveguides, from
the point of view of quadratic forms.  This has consequences for further
conjectures on spectral optimization and for the study of Schr\"odinger
operators depending on curvature.  In the final section we prove a new
theorem on spectral optimization for some one-dimensional Schr\"odinger
operators.

\medskip
%%%%%%%%%%%%%%%%%%%%%%%%%%%%%%%%%%%%%%%
\noindent{\bf {II.  Isoperimetric spectral theorems for annuli and spherical
shells}}
%%%%%%%%%%%%%%%%%%%%%%%%%%%%%%%%%%%%%%%
\medskip

In sections II--III we consider annular domains $D \subset R^{\nu + 1}$,
consisting of
the points on one side of a closed, sufficiently smooth
non--self--intersecting
subset $\Omega$ of
dimension $\nu$, and within a distance $d$ of $\Omega$.  Our theorems will
apply when $\nu = 1$ or $2$, and where $d$ is sufficiently small, thus
corresponding to physical structures such as quantum wires which form
closed loops, or thin capacitors or resonating cavities with special
geometries.  We shall refer to $\Omega$ as the {\it inner edge} or the
{\it outer edge} of the domain.  The area (or length) of
$\Omega$ will be written as $|\Omega|$.

The edge will be assumed sufficiently smooth that its principal
curvatures are defined and bounded at all points (i.e.,
$\partial\Omega \in C^2$), and it is restricted so that all
principal curvatures are bounded
in magnitude
by $1/d$.
(We choose the convention for plane curves which allows the curvature
to have either sign.  Later, when we treat surfaces, they will be assumed
convex, so the principal curvatures will be positive.)
This is an important
constraint, which will be assumed throughout the article.  In the language
of differential geometry, it allows the existence of a Fermi coordinate
system for $D$, consisting of a globally defined coordinate
$r := {\rm distance\  from}\  \Omega$, which is orthogonal to the coordinates on the
smooth ``level surfaces''
$\Omega_r := \left\{x \in D: dist(x,\Omega) = r\right\}$.
(We shall not make special assumptions about the coordinates on the level
surfaces.)  A set $D$ satisfying these assumptions will be called a
{\it smooth annular domain}, and $d$ will be its {\it thickness}.
\smallskip
{\bf Theorem 1:} {\it
a)  With  $dim(\Omega) = 1$, fix the length  $|\Omega|$ and the thickness
$d$,and consider the Dirichlet problem for the Laplacian on all smooth
annular domains with $\Omega$ as one of the edges, either
inner or outer.
Then the
fundamental eigenvalue $\lambda_1$ is uniquely maximized when
$\Omega$ is a circle.

b)  With  $dim(\Omega) = 2$, fix the surface area  $|\Omega|$ and
the volume, and consider the Dirichlet problem for the Laplacian
on all smooth annular domains $D$ with convex outer edges
$\Omega$. Then the fundamental eigenvalue $\lambda_1$ is uniquely
maximized when $\Omega$ is a sphere.}
\smallskip

{\it Remark.}  The assumptions in part a) are tantamount to fixing
the area of the domain $D$, and thus when $\Omega$ is the outer edge
statements a) and b) are analogous.  The claim when $\Omega$
is the inner edge (or, by nearly the same proof, a central level curve)
is an additional fact which appears to be valid only when $\Omega$ is
one--dimensional.

{\it Proof.} a)  We first consider the case $\nu = 1$, and
normalize so that $|\Omega| = 2  \pi$. By the Rayleigh principle,
$$ {\lambda }_{1}\ =\ \inf\int\!\!\!\int_{}^{}{\left|{\nabla \zeta
}\right|}^{2}\ {\d}^{2}x, $$ where the infimum is taken over
smooth functions $\zeta$ on the closure of $D$ which vanish on its
boundary, normalized in $L^2$. We write this in the orthogonal
coordinate system defined by $r$ and $s:=$ arclength of $\Omega$
of the point nearest to ${\bf x}$, as measured counterclockwise
from some reference position:
$$ {\lambda }_{1}\ =\ \inf\int_{0}^{d}\int_{0}^{2\pi }\left({{1
\over \left({1 \pm \kappa (s)\ r}\right)}{\zeta }_{s}^{2}+\
\left({1 \pm \kappa (s)\ r}\right){\zeta }_{r}^{2}}\right)\ \d s\
\d r $$ (e.g., see [Ban], p. 143). Here, $\kappa$ is the curvature
of $\Omega$ at $s$, and we have chosen the orientation whereby the
plus sign corresponds to $\Omega$ being the inner edge, and the
minus sign to it being the outer edge. Suppose now that a smooth
test function $\zeta$ which vanishes on the boundary of $D$ is
independent of $s$. Then

$$
{\lambda }_{1}\ \le \ \int_{0}^{d}\int_{0}^{2\pi }\left({\
\left({1 \pm \kappa (s)\ r}\right){\zeta }_{r}^{2}}\right)\ \d s\ \d r\ =\
\int_{0}^{d}\left({\ 2 \pi \left({1 \pm r}\right){\zeta }_{r}^{2}}\right)\
\d r.
$$

The inequality would be strict if the true ground state were to
depend on $s$.  The final expression, however, is equivalent to
the one for the corresponding annulus, restricted to the set of
test functions independent of $s$.  Since the fundamental
eigenfunction for the circular annulus is attained in the set of
functions independent of $s$, we conclude

$$
{\lambda }_{1}\ \le \rm \ {\lambda }_{1}\left({annulus}\right).
$$
Because equality requires that the fundamental eigenfunction be
independent of $s$, one of the two terms in the eigenvalue
equation which includes $\kappa(s)$ is zero due to the presence of
$\zeta_s$. This forces the only remaining term to be independent
of $s$ for any curve $\Omega$ which maximizes $\lambda_1$.  The
only possibility is a circle.

b)  Next we turn to the case $\nu = 2$, and normalize so that
$|\Omega| = 4 \pi$, as for the unit sphere.  We begin as before, by using
the coordinate $r$, supplemented by coordinates orthogonal to $r$ on the
level surfaces $\Omega_r$.  It is convenient to denote $A(r)
:= |\Omega_r|$.
Assuming that the test functions depend only on $r$, the Rayleigh
principle states (for normalized test functions):

$$
{\lambda }_{1}\ \le  \ \int_{0}^{d}{\zeta }_{r}^{2}\ A(r) \d r.
$$
We recall here that now the thickness $d$ depends on the domain $D$;
specifically, they are connected by

$$
\int_{0}^{d}A(r) \d r\ =\ Vol(D).
$$

Let us now change variables to $r'$ defined so that
      $$A_0(r') \d r' = A(r) \d r,$$
where $A_0(r') := 4 \pi (1 - r')^2 =$ the area of the sphere of radius
$1 - r'$.  We find:
$$
{\lambda }_{1}\ \le \ \int_{0}^{d'}{\zeta }_{r'}^{2}\ {\left({{A(r)
\over {A}_{0}(r')}}\right)}^{2}{A}_{0}(r')\d r',
$$
where $d'$ is the thickness of the spherical shell with the same volume as $D$.
(For brevity we use informal notation for functions of transformed
variables, etc.)

We next claim that the expression in parentheses is strictly
smaller than $1$ for $r > 0$, unless $\Omega$ is a sphere.  This
is because the growth rate of the volume (here, area) element of
the level surfaces is the sum of the principal curvatures (e.g.,
[Kar], eq. (1.5.4) or [Spi]. p. 418), and hence
$$ A'(r)\ =\ -\
\int_{{\Omega }_{r}}^{}\left({{\kappa }_{1}+{\kappa
}_{2}}\right){\d}\sigma \ \eqno(2.1)
$$
(cf. [Spi], p. 426). The negative sign arises here because
$\Omega$ is the outer edge of the domain. At this stage we remark
that if $\Omega$ is convex, then so are all the level surfaces in
a smooth annular domain bounded by $\Omega$, and equation (2.1)
can be integrated in a closed form discovered by Steiner ([San],
p. 325):
$$ A(r) \ =\ A(0)\ -\ r\ \int_{{\Omega}_{0}}^{}\left( {\kappa_1+
\kappa_2} \right) \ d\sigma +\ 4\pi r^2 $$
Now recall a classical geometric lemma (e.g., [HiCo], p. 225):
\smallskip
{\it Among all closed surfaces of fixed area, bounding a convex
set, the sphere is the unique minimizer of  the total mean
curvature, $\int_{{\Omega }_{r}}^{}\left({{\kappa }_{1}+{\kappa
}_{2}}\right){\d}\sigma$. }
\smallskip
{\noindent We see that for a convex surface of a given area, the
sphere is the unique maximizer of the negative quantity $A'(r)$ for
$r < d$. It then also
follows that on this interval, $r' \le r$ and
$0 < A(r) \le A_0(r) \le A_0(r')$ with equality only in the case of the
sphere. }

The argument then proceeds as in part a):  The upper bound obtained is equal
to the Rayleigh quotient for the spherical shell, with radial test functions.
Since the lowest eigenfunction for the spherical shell is radial, ${\lambda
}_{1}\ \le \rm \ {\lambda }_{1}(shell)$, as claimed.\hfill\lanbox

\smallskip

Theorem 1 contrasts with the Faber--Krahn theorem,
which might lead one to expect that round annular domains were
minimizers rather than maximizers.  There are indeed other situations
where $\lambda_1$ is maximized for the
Laplace operator when the
volume is fixed.  One of these is addressed in a theorem of
Payne and Weinberger [PaWe], on annular domains of a given area,
where the outer edge is subject to Dirichlet conditions and the inner edge to
Neumann conditions.
(For some extensions of this theorem, see [Ban], section 4.3.)

Moreover, for the pure Dirichlet problem, the following is a special case of a
theorem in  [HaKrKu]:
\smallskip
{\it Let the ball $B_1$ be contained inside the ball $B_2$.  Fix
the radii of $B_{1,2}$ but not the position of $B_1$ in $B_2$.
Among all domains of the form $B_2\setminus B_1$, the one with the
highest fundamental Dirichlet eigenvalue for the Laplacian is the
one where they are concentric. }
\smallskip

In these cases as well as in Theorem 1, the domain is allowed to vary
within a class of non-simply connected domains.  The contrast with
Faber--Krahn thus seems to be a topological effect.

For reasons to be explored in the final section, we conjecture that the
conclusion of the theorem is false if the dimension is increased to $\nu > 2$.

\medskip
%%%%%%%%%%%%%%%%%%%%%%%%%%%%%%%%%%%%%%%
\noindent{\bf {III.  The natural coordinate transformation for the Dirichlet quadratic form}}
%%%%%%%%%%%%%%%%%%%%%%%%%%%%%%%%%%%%%%%
\medskip

In this section we begin with a coordinate transformation that is old and
quite standard
(e.g., [daC], [DuEx]), using the Fermi coordinate system of the
previous section.  Our excuse for thus risking the reader's boredom is that
we intend to use the transformation to illuminate an essential geometric
feature of the effective potential--energy term which emerges.  Also, we
shall do the transformation in the context of quadratic forms rather than
operators, which appears to us more direct and offers the possibility of
relaxing some conditions of smoothness.

We continue to assume that our domain is a smooth annular domain with
inner or outer edge $\Omega$.  We construct a Fermi coordinate system
using $r$, the distance from $\Omega$, as one coordinate, orthogonal to
the rest.  We denote the components of the gradient which are parallel to
the level surfaces of constant r by ${\nabla }_{||}\ f.\ \ $

We split the components of the Dirichlet form for the Laplace operator as:
$$
\int_{D}^{}{\left|{{\nabla }_{||}\ \zeta }\right|}^{2} \  {\d}^{\nu \rm +1}x\
+\ \int_{D}^{}{\left|{{\zeta }_{r}}\right|}^{2}\ {\d}^{\nu \rm +1}x\
$$
and transform the second term (only) as follows.  Fix a coordinate system
on the edge $\Omega$, and for any point {\bf x} in $D$, choose as its
coordinates other than r the coordinates of the closest point on the edge.
Let $dV^{\nu}$ denote the volume element on $\Omega$.  Then
$$
\int_{D}^{}{\left|{{\zeta }_{r}}\right|}^{2}\ {\d}^{\nu +1}x\ =\ \int_{\Omega
}^{}\ \int_{0}^{d}\ {\left|{{\zeta }_{r}}\right|}^{2}\rho \rm (\bf x\rm ) \d r\
\d {V}^{\nu },
$$
where $\rho$ is the {\it volume growth factor}, a familiar quantity in
differential geometry [Kar] [Spi].  We write the test function as
$$
\zeta \rm \ =\ {1 \over \sqrt {\rho }}\cdot \rm \ \left({\sqrt {\rho }\
\zeta }\right)
$$
and use the product rule in the form
$${\left({\left({fg}\right)'}\right)}^{2}\ =\ {f}^{2}\
{\left({g'}\right)}^{2}\ +\ {g}^{2}\ {\left({f'}\right)}^{2}\ +\ {1 \over 2}\
{\left({{f}^{2}}\right)}^{\prime}{\left({{g}^{2}}\right)}^{\prime}
$$
to find

$$
\int_{0}^{d}\ {\left|{{\zeta }_{r}}\right|}^{2}\rho \rm \d r\ =\
\int_{0}^{d}\left({{\left|{{\left({\sqrt {\rho }\zeta
}\right)}_{r}}\right|}^{2}\ +\ {1 \over 4}\ {\left({{{\rho }_{r} \over \rho
}}\right)}^{2}{\zeta }^{2}\rho \rm \ +\ {\rho  \over \rm 2}{\left({{1 \over
\rho }}\right)}_{r}{\left({\rho {\zeta }^{\rm 2}}\right)}_{r}}\right)\d r.
$$

When the final term is integrated by parts, the full Dirichlet form
takes on the appearance
$$
\int_{A}^{}{\left|{{\nabla }_{||}\ \zeta }\right|}^{2}\ {\d}^{\nu \rm +1}x\
+\ \int_{A}^{}q(\bf x\rm )\ {\zeta }^{2}\ {\d}^{\nu \rm +1}x\ +\ \ \int_{\Omega
}^{}\ \int_{0}^{d}{\left|{{\left({\sqrt {\rho }\zeta
}\right)}_{r}}\right|}^{2}\ \d r\  \d {V}^{\nu }. \eqno(3.1)
$$
The effective potential in the middle contribution is
$$
q(\bf x\rm )\ :=\ -{1 \over 4}{\left({{{\rho }_{r} \over \rho
}}\right)}^{2}+\ {1 \over 2}{{\rho }_{rr} \over \rho }.
$$

In order to elucidate the geometric meaning of the effective potential,
we recall two elementary facts about the geometry of level surfaces:
$$
\eqalign{(i) &\  {\partial  \over \partial  r}\rho  \ =\ \pm
\left(\sum\limits_{ j\ =\ 1}^{\nu } {\kappa }_{ j}\right) \rho  \cr
(ii)&\ {\partial  \over \partial r}
\sum\limits_{j\ =\ 1}^{\nu } {\kappa }_{ j} =\
\mp \sum\limits_{j\ =\ 1}^{\nu} {\kappa }_{j}^{2}}.$$
Here, $\kappa_j$ are the principal curvatures {\it of the level surface at
coordinate r},
and the choice of sign corresponds as usual to the choice of orientation.
The signs of (i) and (ii) anticorrelate.
(The two-dimensional version of (i) made an appearance
in the previous section.)

A short calculation reveals an identity
with unambiguous sign regardless of orientation:
\smallskip
{\bf Proposition 2:} {\it The Dirichlet form for the Laplacian is
given by (3.1) with
$$
q(\bf x\rm )\ =\ {1 \over 4}{\left({\sum\limits_{\rm j\ =\ 1}^{\nu } {\kappa
}_{\rm j}}\right)}^{2}\ -\ {1 \over 2}\ \sum\limits_{\rm j\ =\ 1}^{\nu }
{\kappa }_{\rm j}^{\rm 2}. \eqno (3.2)
$$
}
\smallskip

Effective potentials of the form (3.2) have long been familiar in
the study of microstructures with $\nu = 1$ or $2$ [daC] [DuEx]
[Ex\v Se]. For the most part they have heretofore appeared as
asymptotic calculations with the principal curvatures calculated
on the edge, rather than on the level surfaces.  In [DuEx] and
[Ex\v Se] a variant effective potential throughout the domain is
obtained by a ``straightening'' transformation.  Roughly speaking,
the latter articles replace the first term in (3.1) by an integral
on a flat manifold, thus avoiding metric tensors, at the price of
some derivatives of curvature in the potential energy.

We content ourselves here with a few simple observations about the uses of
Proposition 2, which we hope to explore further in a future article.

First, Proposition 2 is a tool for obtaining spectral bounds.  A
representative such bound is:
\smallskip
{\bf Corollary 3:} {\it
$$
{\lambda }_{1}\ \ge \rm \ {{\pi }^{2} \over {d}^{2}}\ +\ \inf(q(\bf x\rm
)).
$$
}
\smallskip
{\it Proof.} This follows immediately once it is realized that the
final term in (3.1) can  be considered as the Rayleigh quotient
for the simple operator $-d^2/dx^2$  with Dirichlet boundary
conditions at $0$ and $d$, because the normalization is unchanged:
${\left|\!\left|\sqrt {\rho }\zeta \right|\!\right|
}_{{L}^{2}(D,\d r \d {V}^{\nu })}\ =\ {\left|\!\left|\zeta
\right|\!\right| }_{{L}^{2}(D,\ {\d}^{\nu \rm +1}x)}\ $.

The first term of (3.1) is dropped and in the second the potential is
replaced by its infimum.\hfill\lanbox
\smallskip

Actually, sharper bounds than Corollary 3 are obtainable using analogues of
the Boggio and Hardy inequalities [Bog] [Dav].

Next we observe that although the effective potential is nonpositive if
$\nu = 1$ ($q(\bf x\rm )\ =\ -{{\kappa }^{2} \over 4}\ $)
or $2$ ($q(\bf x\rm )\ =\ -{{\left({{\kappa }_{1}-{\kappa }_{2}}\right)}^{2}
\over 4}\ $),
as soon as  $\nu > 2$ the effective potential becomes positive for the sphere
and many other hypersurfaces.  While we have neither an
isoperimetric theorem nor a counterexample to offer when
$\nu > 2$, the simplified model in the next section
indicates that the situation can
change dramatically when the effective potential may be positive.

\medskip
%%%%%%%%%%%%%%%%%%%%%%%%%%%%%%%%%%%%%%%
\noindent{\bf IV.  Spectral optimization for some Schr\"odinger operators
depending on curvature}
%%%%%%%%%%%%%%%%%%%%%%%%%%%%%%%%%%%%%%%
\medskip

In order to begin the analysis of the case where the effective potential is
quadratic in the curvature but potentially positive, we consider the lowest
eigenvalue of a family of one-dimensional operators parametrized by a
real coupling constant $g$,
$$
H(g)\ :=\ -{{d}^{2} \over {ds}^{2}}\ +\ g\ {\kappa }^{2}.
$$

We have thus simplified the situation of the previous section by
reducing the dimension, but have allowed the effective potential
to be positive when $g > 0$. This model is also a natural
generalization of one examined in [HaLo], corresponding to the
case $g = -1$.

The operator $H(g)$ is defined on a closed planar curve normalized
to have length $1$. If $g < 0$, it is straightforward to see that
of all curves, the circle maximizes the fundamental eigenvalue,
as one might conjecture from Theorem
$1$ by considering infinitesimally thin annular domains
approximating a closed curve.
This would
correspond only to $g=-1/4$, of course, but for all $g < 0$ the statement is
easy to obtain from the Rayleigh--Ritz inequality by considering the trial
function $\zeta = 1$  and making a simple estimate with the
Cauchy--Schwarz inequality.

The situation is more subtle when $g > 0$.
\smallskip
{\bf Theorem 4:} {\it a)  Suppose that  $0 < g < 1/4$.
Then the circle is the unique
curve which  minimizes the fundamental eigenvalue  $\lambda_1$.

b)  Suppose that  $1 < g$.  Then the circle does not minimize the
fundamental eigenvalue  $\lambda_1$.}
\smallskip

{\it Remarks.} The questions of the critical value of $g$ and the nature of the
transition remain open.  We conjecture that there is no legitimate
minimizing curve when $g > 1$.

{\it Proof.} a)  Assume first that $0 < g < 1/4$.  The minimal value of $\lambda_1$,
which  we denote $\lambda_*$, is

$$
\matrix{\inf\cr
\kappa \cr}\matrix{\inf\cr
\zeta \cr}\ \int_{}^{}\left({{\left({{\d \zeta  \over \d s}}\right)}^{2}\ +\
g\ {\kappa }^{2}{\zeta }^{2}}\right) \d s,
$$

{\noindent where the normalized  $\zeta$ varies over the class of
smooth periodic positive functions, while
$\kappa$ is the curvature function defining a planar curve.  The
assumption of positivity may be imposed because of the positivity property
of ground states of Schr\"odinger operators.  We relax
the conditions on $\kappa$ so that we require only that it be a function
with integral $2 \pi$, i.e., we do not explicitly require that the curve be
closed.  We shall see that the minimizing curve is still a circle and
therefore closed, so this causes no harm.
}

Because the quantity in question is an iterated infimum, it may be
calculated in the other order. By Cauchy-Schwarz's inequality
$$
2 \pi = \int{\kappa \over \zeta} \zeta \d s \leq \left( \int {1 \over \zeta^2}
\d s \right)^{1/2} \left( \int \kappa^2 \zeta^2 \d s \right)^{1/2},
$$
with equality only if
$$
\kappa = \left( 2\pi / \int {1 \over \zeta^2} \d s\right){1 \over \zeta^2 }.
$$
Hence $\lambda_*$ is identical to the infimum of the
unusual functional

$$
E \left({\zeta }\right) \ :=\ \ \int_{}^{}{\left({\d \zeta  \over
\d s}\right)}^{2} \d s\ +\ {4{\pi }^{2}g \over \int_{}^{}\left({\ {1 \over {\zeta
}^{2}}}\right) \d s}.
$$
By choosing the trial function which becomes exact for the circle,
i.e., $\zeta \equiv 1$, it follows that $\lambda_*\le E(\zeta) =
4g \pi^2 < \pi^2$ for $g<1/4$.

In order to establish the existence of a minimizer for
$E\left({\zeta }\right)$, we need a pointwise estimate:
\smallskip
{\bf Lemma 5:} {\it If  $E(\zeta) \le \pi^2$  for a positive test function  $\zeta$
normalized in $L^2$, then
$$ {\inf}_{s}\left({\zeta \left({\rm s}\right)}\right)\ > \ 1\ -\
{\sqrt {E\left({\zeta }\right)} \over \pi }. $$
}
\smallskip
{\it Proof of Lemma 5.}
$$ E\left({\zeta }\right)\ >\
\int_{0}^{1}{\left({\zeta^\prime}\right)}^{2}\ \d s\ =\
\int_{0}^{1}{{\left({\zeta \rm \ -\ {\zeta
}_{\min}}\right)}^{\prime}}^{2}\d s\ \geq \ {\pi
}^{2}\int_{0}^{1}{\left({\zeta \rm \ -\ {\zeta
}_{\min}}\right)}^{2}\d s, $$
because $\zeta - \zeta_{min}$ is an admissible test function for the operator
$-d^2/ds^2$ with an additional Dirichlet boundary condition at the position
of $\zeta_{min}$, and that operator is bounded below by
its lowest eigenvalue $\pi^2$,
in the sense of quadratic forms.  Hence
$$
E\left({\zeta }\right)\ >\ \ {\pi }^{2}\left({\int_{0}^{1}{\zeta }^{2}\d s\
-\ 2\ {\zeta }_{\min}\int_{0}^{1}\zeta \rm \d s\ +\ {\zeta
}_{\min}^{2}}\right),
$$
which implies the claimed bound when we use the Cauchy-Schwarz
inequality to replace $\int_{0}^{1}\zeta \rm \ ds\ \le \rm \ \sqrt
{\int_{0}^{1}{\zeta }^{2}\d s}=1$  and then solve for
$\zeta_{min}$. \hfill\lanbox

We now continue with the proof of Theorem 4.  Because of the lemma,
if $\lambda_* < \pi^2$, then any minimizing sequence for
$E\left({\zeta }\right)$ is bounded in the Sobolev space ${H}_{per}^{1}$.
By a standard compactness theorem, a subsequence converges uniformly to
a limit $\zeta_*$ which, by the lemma, is strictly positive.  The same
function is also a weak limit in the ${H}_{per}^{1}$ sense of a
subsequence, from which it follows that
$E(\zeta_*) = \lambda_*$ as claimed.

We have thus established the existence of the minimizer $\zeta_*$,
which is a nonnegative periodic function on the interval $[0,1]$.
Next we observe that $\zeta_*$ satisfies the Euler equation for
the functional $E$, which is found to be
$$-{\zeta }_{{}^*}^{''}\ +\ M\ {1 \over {\zeta }_{{}^*}^{3}}\ =\
C\ {\zeta }_{{}^*}, \eqno(4.1) $$
where
$$ M\ =\ {4{\pi }^{2}g \over {\left({\int_{0}^{1}{1 \over {\zeta
}_{{}^*}^{2}}\ ds}\right)}^{2}} $$
and $C$ is a Lagrange multiplier. Since $E(\zeta_*)\ = \lambda_*$,
we find $C\ =\lambda^*$ by multiplying (4.1) by $\zeta_*$ and
integrating by parts.

If the minimizer is a constant function, $\zeta_*(s) =1$, we have
$\lambda_*\ =\ M\ = 4 \pi^2 g$. It remains to be seen whether
$\zeta_*$ may be nonconstant. We multiply equation (4.1) by
$\zeta_*^{\prime}$ and integrate; this yields
$$ {{\zeta_* }^{\prime}}^2\ +\ {M \over \zeta_*^2} +\ \lambda_*
\zeta_*^2\ =\ C' \eqno(4.2)$$
for some $C'$. The minimizer is normalized by assumption, so
integrating (4.2) we find $C'\ =\ 2\lambda_*$. Hence the last
equation can be rewritten as
$$ {{\zeta_* }}^2 {{\zeta_* }^{\prime}}^2\ =\ \lambda_*\ -\ M\ -
\lambda_* \left( {{\zeta_* }}^2 - 1 \right)^2. $$
It follows that $\zeta_*$ oscillates between its (positive)
minimal value $\sqrt{1- \sqrt{1-M/\lambda_*}}$ and the maximum
$\sqrt{1+ \sqrt{1-M/\lambda_*}}$, being strictly monotonic between
them. The corresponding solution of (4.1) is given by

$$ {\zeta }_{{}^*}^{2}\ =\ 1 + \sqrt{1-M/ \lambda_{{}^*}} \
\cos\left({2 \sqrt{\lambda_{{}^*}} (s-s_0)}\right); $$
it is unique if we fix the point $s_0$ where the maximum is
reached. However, a nonconstant function of this type cannot be
periodic on $[0,1]$, because $\lambda_*\ <\ \pi^2$.

b)  Suppose now that $g > 1$.  The lowest eigenvalue for the circle is
$4 \pi^2 g$, so we need to show that lower
eigenvalues are attainable,
which we proceed to do with an explicit example.
(While our example will have a discontinuous curvature, it can
be approximated arbitrarily well by curves for which $\kappa(s)$ is perturbed  arbitrarily slightly in the $L^{2}$ sense,
which implies that the eigenvalues are shifted by arbitrarily
small amounts [ReSi].)
Consider the thin
stadium--shaped curve defined by
$\kappa \left({\rm s}\right)\rm \ =\ {\pi  \over \varepsilon }\ \ $
for $1/2-\epsilon < s < 1/2$ and $1-\epsilon < s < 1$, and 0 otherwise,  for
some $\epsilon << 1/2$.  Now estimate the fundamental eigenvalue using
the Rayleigh-Ritz inequality and the trial function
$\sin\ \left({{\pi \rm \ s \over {1 \over 2}\ -\ \varepsilon }}\right)$
for $0 < s < 1/2-\epsilon$ and $0$ otherwise;
the result is
${{\lambda }_{1}\ \le \rm \ \left({{\pi  \over {\rm 1 \over 2}\rm \ -\
\varepsilon }}\right)}^{2},\ \ $
which can be made arbitrarily close to $4 \pi^2$.\hfill\lanbox

{\it Remark:} The proof of a) shows that the circle is also a minimizer
if $g=1/4$, however we do not know whether it is the unique minimizer.

\medskip
%%%%%%%%%%%%%%%%%%%%%%%%%%%%%%%%%%%%%%%
\noindent{\bf {Acknowledgments}}
%%%%%%%%%%%%%%%%%%%%%%%%%%%%%%%%%%%%%%%
\medskip

The authors are grateful to Mark Ashbaugh, Michiel van den Berg,
Rick Laugesen, and the referee for comments and references. E.H.
wishes to thank the Erwin Schr\"odinger Institut and the Nuclear
Physics Institute at \v{R}e\v{z} u Prahy for hospitality while
some of this work was done.

\medskip
%%%%%%%%%%%%%%%%%%%%%%%%%%%%%%%%%%%%%%%
\noindent{\bf {References}}
%%%%%%%%%%%%%%%%%%%%%%%%%%%%%%%%%%%%%%%
\medskip

\item{[AlFu]}  Nicholas D. Alikakos and Giorgio Fusco,  The spectrum of the
Cahn-Hilliard operator for generic interface in higher space dimensions,
{\it Indiana U.  Math. J.} {\bf 4}, 1993, pp. 637--674.

\item{[AsBe]}  Mark S. Ashbaugh and Rafael D. Benguria,
Isoperimetric inequalities for eigenvalue ratios, pp. 1--36 in:
A. Alvino, E. Fabes, and G. Talenti, eds.,
Partial Differential Equations of Elliptic Type, Cortona,
1992.
Cambridge:  Cambridge University Press, 1994.

\item{[Ban]}  Catherine Bandle,
Isoperimetric inequalities and applications,
Pitman Monographs and Studies in Mathematics {\bf 7}.  Boston:  Pitman, 1980.

\item{[Bog]}  Tommaso Boggio, Sull'equazione del moto vibratorio delle
membrane elastiche,  {\it Rend. Accad. Lincei}, sci. fis., ser. 5 {\bf 16}(1907)386-393.

\item{[daC]}  R.C.T. da Costa, Quantum mechanics of a constrained
particle,
Phys Rev. {\bf A23}(1981) 1982--1987 (and later articles).

\item{[Dav]} E.B. Davies, A review of Hardy inequalities, preprint 1998
available electronically as
http://xxx.lanl.gov/abs/math.SP/9809159.

\item{[DuEx]}  Pierre Duclos and Pavel Exner,
Curvature--induced bound states in quantum wave\-guides in two and
three dimensions, {\it Rev. Math. Phys.} {\bf 7}(1995)73--102.

\item{[Ex\v Se]}  Pavel Exner and Pavel \v{S}eba, Bound states
in curved quantum waveguides, {\it J. Math. Phys.}
{\bf 30}(1989)2574--2580.

\item{[Fab]}  G. Faber, Beweis, dass unter allen homogenen
Membranen von gleicher Fl\"ache und gleicher Spannung
die kreisf\"ormige den tiefsten Grundton gibt,
{\it Sitzungsber. der mathematisch-physikalischen Klasse der Bayer.
Akad. der Wiss. zu M\"unchen} (1923)\hfill 169--172.

\item{[Har]} Evans M. Harrell II,
On the second eigenvalue of the Laplace operator penalized by curvature,
{\it Journal of Differential Geometry and Applications}
{\bf 6}(1996)397--400.

\item{[HaKrKu]}  Evans M. Harrell II, Pawel Kr\"oger, and Kazuhiro
Kurata, work in progress.

\item{[HaLo]}  Evans M. Harrell II and Michael Loss,
On the Laplace operator penalized by mean curvature, {\it Commun.
Math. Physics} {\bf 195}(1998)643-650.

\item{[Her1]} Joseph Hersch, Quatre propri\'et\'es isop\'erimetriques
de membranes sph\'eriques \hfill\break homog\`enes,
{\it C.R. Acad. Sci. Paris}, s\'er A-B {\bf 270}, 1970, pp. A1645--1648.

\item{[Her2]}  Joseph Hersch, Isoperimetric monotonicity:
Some properties and conjectures (connections between isoperimetric
inequalities), {\it SIAM Review} {\bf 30}(1988)551--577.

\item{[HiCo]}  D. Hilbert \& S. Cohn--Vossen.
Geometry and the Imagination.  New York:
Chelsea, 1952.

\item{[Kar]}  Hermann Karcher, Riemannian comparison constructions,
pp. 170-222 in: S. Chern, editor, Global Differential Geometry,
Studies in Mathematics {\bf 27}.  Washington:  Math. Assoc. Amer.,
1989.

\item{[Kra]}  E. Krahn, \"Uber Minimaleigenschaften der Kugel in
drei und mehr Dimensionen,
{\it Acta Comm. Univ. Tartu} (Dorpat) {\bf A9}(1926)1--44.

\item{[PaWe]}  L.E. Payne and H.F. Weinberger, Some isoperimetric
inequalities for membrane frequencies and torsional rigidity,
{\it J. Math. Anal. Appl.} {\bf 2}(1961)210--216.

\item{[ReSi]}  M. Reed  and B.Simon,
Methods of Modern Mathematical Physics, {\bf IV}:
Analysis of Operators. New York: Academic Press, 1978.

\item{[Spi]}  Michael Spivak,
A Comprehensive Introduction to Differential Geometry, {\bf IV},
second edition.
Houston:  Publish or Perish, 1979.

\item{[San]} L.A. Santalo,
Integral Geometry, pp.303--350, in: S. Chern, ed., {\it Global
Differential Geometry}, Studies in Mathematics {\bf 27}.
Washington: Math. Assoc. Amer., 1989.

\item{[Wei]} H.F. Weinberger, An isoperimetric inequality for the
n--dimensional free membrane problem,
{\it J. Rat. Mech. Anal.} {\bf 5}(1956)633--636.

\end